\documentstyle[amssymb,aps,prb,twocolumn,epsfig]{revtex}

\draft

\begin{document}

\twocolumn[\hsize\textwidth\columnwidth\hsize\csname@twocolumnfalse\endcsname

\title{ Kondo impurities in nanoscopic systems: new confinement-induced regimes}

\author{P. S. Cornaglia and C. A. Balseiro}

\address{Instituto Balseiro and Centro At\'omico Bariloche, Comisi\'on
Nacional de Energia At\'omica, 8400 San Carlos de Bariloche, Argentina.}

\date{\today}
\maketitle

\begin{abstract}
We present results for Kondo impurities in nanoscopic systems. Using exact diagonalization in small clusters and Wilson's renormalization group we analyze an isolated system and a nanoscopic system weakly coupled to a macroscopic reservoir. In the latter case, new regimes not observed in macroscopic homogeneous systems are induced by the confinement of conduction electrons. These new confinement-induced regimes are very different depending on whether the Fermi energy coincides with the energy of a resonant state or lies between two quasi-bound states. 
\end{abstract}

]

The physics of nanoscopic and mesoscopic systems became of grate interest
due to the possibility of making experiments in extremely small samples.
Some examples of nano and mesoscopic systems recently studied are small
metallic and superconducting islands,\cite{sup} quantum dots,\cite{QD}
nanotubes,\cite{nanotu} and quantum corrals.\cite{Mano} Depending on the
type of system under study, different parameters can be controlled.

The advances in nanotechnologies revived the interest in the Kondo effect,%
\cite{Kondo} one of the paradigms of strongly correlated systems. On one
hand, Scanning Tunneling Microscopy (STM) allowed the direct measurement of
local spectroscopic properties of Kondo impurities.\cite{Mano,stm} On the
other hand it has been shown that quantum dots and single walled carbon
nanotubes weakly coupled to contacts may behave as Kondo impurities
generating new alternatives to study the phenomena.\cite{Lee}

In this work we report results for Kondo impurities in nanoscopic systems.
Some experiments show that the effects of confinement are important and
recently Affleck and Simon\cite{Affleck} proposed to use a mesoscopic system
to measure the Kondo screening length. Although the problem has been
addressed theoretically,\cite{sb,HCB,aligia} there are still many open
questions.\cite{ps} What is the temperature dependence of the Kondo
screening in nanoscopic systems? What is the effect adding leads or coupling
the system to a macroscopic reservoir?

In the case of systems weakly coupled to a macroscopic reservoir, the one
electron states acquire a finite lifetime and the spectrum consists of
resonant states. This situation is the most relevant to compare with
experiments, however due to the intrinsic difficulties to formulate the
problem is has not been analyzed in detail. As an example we may mention the
case of quantum corrals, were the electrons are not well confined in the
corral due to leaking of the wave function and hybridization with the bulk
states. Consequently the number of electrons in the corral is not a good
quantum number and the local density of states as a function of the energy
presents well defined maxima corresponding to the quasi-confined states of
the corral.\cite{mirageps} We show that in systems like the quantum corrals
new regimes are induced by the confinement of conduction electrons. These
new confinement-induced regimes are very different if the Fermi energy is at
resonance or between two quasi-bound states. We also discuss the case of
quantum dots in Aharonov-Bohm rings.

We have performed numerical diagonalization in small clusters and
implemented Wilson's numerical renormalization group (NRG).\cite{Wilson} Our
starting point is the Anderson model for magnetic impurities. The model
Hamiltonian reads:\cite{and}

\begin{eqnarray}
H_{AM} &=&\sum_{\sigma }\varepsilon _{d}d_{\sigma }^{\dagger }d_{\sigma }+%
{\it U}d_{\uparrow }^{\dagger }d_{\uparrow }d_{\downarrow }^{\dagger
}d_{\downarrow }+\sum_{\nu ,\sigma }\varepsilon _{\nu }c_{\nu \sigma
}^{\dagger }c_{\nu \sigma }  \nonumber \\
&&+\sum_{\nu ,\sigma }V_{\nu }c_{\nu \sigma }^{\dagger }d_{\sigma }+V_{\nu
}^{*}d_{\sigma }^{\dagger }c_{\nu \sigma }-\mu _{i}BS_{iz},  \label{ham}
\end{eqnarray}
here the operator $d_{\sigma }^{\dagger }$ creates an electron with spin $%
\sigma $ at the impurity orbital with energy $\varepsilon _{d}$ and Coulomb
repulsion ${\it U}$, $c_{\nu \sigma }^{\dagger }$ creates an electron in an
extended state with quantum numbers $\nu $ and $\sigma $ and energy $%
\varepsilon _{\nu }$. The last term represents the effect of an external
magnetic field along the $z$-direction coupled to the impurity spin ${\bf S}%
_{i}=d_{\sigma }^{\dagger }{\mathbf{\tau} }_{\sigma ,\sigma ^{\prime }}d_{\sigma
^{\prime }}$ where ${\mathbf{\tau} }$ are the Pauli matrices.

We start by analyzing the case of an isolated system. The properties of the
system depend crucially on a few parameters, in particular we show below
that in the Kondo limit a finite system presents quite a marked even-odd
electron number asymmetry.

We first consider the case of Hamiltonian (1) describing a linear chain of $%
N $ equivalent sites with the impurity at one end. In this case $%
\epsilon _{\nu }=-2t\cos (\nu \pi /(N+1))$, where $t$ is the hopping matrix
element in the chain and $V_{\nu }=V_{0}\sin (\nu \pi /(N+1))$ with $\nu
=1,2,\ldots ,N$. We performed exact diagonalization in these finite systems
using a Lanczos algorithm.

\begin{figure}[t]
\epsfxsize = 6.5cm
\begin{center}
\leavevmode
\epsffile{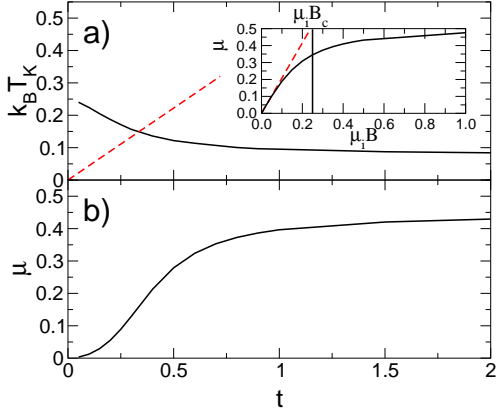}
\end{center}
\caption{(a) Kondo temperature, as defined in the text, versus hopping
integral $t$ for a chain of nine sites with eight particles ($U=1.0$, $%
V_{0}=0.3$, and $\varepsilon _{d}=\varepsilon _{F}-0.5$). The dashed line is
the level spacing $\Delta $. Inset: impurity magnetization versus $B$. (b)
Expectation value of the impurity spin versus $t$, for seven electrons, and
the same parameters as in (a).}
\label{fig1}
\end{figure}

For an even number of electrons in the system, the ground state is a singlet
indicated as $|0\rangle $ and the expectation value of the impurity spin $%
\langle 0|S_{iz}|0\rangle $ is zero. The impurity spin is screened by the
free electrons and the zero temperature susceptibility is finite. We can
estimate the characteristic energy scale $k_{B}T_{K}$ as the energy involved
in the screening of the impurity spin. The magnetization as a function of
the external field $B$, shown in the inset of Fig.\ref{fig1}(a), gives an
estimation of $T_{K}$. We define $k_{B}T_{K}=\mu _{i}B_{c}$ where $B_{c}$ is
the crossover field indicated in the inset. As $t$ increases the mean
density of conduction states decreases and consequently $T_{K}$ decreases.
As indicated in Fig.\ref{fig1}(a) two regimes are obtained: $%
k_{B}T_{K}<\Delta $ and $k_{B}T_{K}$ $>\Delta $ where $\Delta \simeq 4t/N$
is the level spacing. For the parameters used the crossover between these
two regimes takes place at a hopping $t_{cross}\simeq 0.27$. 
For an odd number of electrons, the ground state is a Kramers spin-$1/2$
doublet, $|\Uparrow \rangle $ and $|\Downarrow \rangle $. The expectation
value $s\equiv \langle \Uparrow |S_{iz}|\Uparrow \rangle =-\langle
\Downarrow |S_{iz}|\Downarrow \rangle $ is different from zero and, in the
low temperature limit the impurity susceptibility diverges as $\chi =\mu
_{i}^{2}s^{2}/k_{B}T$. In Fig.\ref{fig1}(b) the effective magnetic moment $%
\mu =\mu _{i}s$ (we take $\mu_i=1$) is plotted as function of $t$. For large $\Delta $ (large 
$t$) the impurity spin is unscreened and $s\approx 1/2$ as expected for a
gaped system\cite{Bu}; for small $\Delta $ the impurity spin tends to be
completely screened. The crossover between these two regimes takes place at $%
t_{cross}$.

Although these results were obtained for a very small system, they give the
correct physical picture for zero temperature. While for $\Delta \ll
k_{B}T_{K}$ the impurity spin is essentially screened regardless of the
parity of the electron number, for $\Delta \gtrsim k_{B}T_{K}$ an important
even--odd asymmetry is obtained. In order to analyze larger systems we have
to resort to some approximate scheme. We have adapted the NRG to the present
problem. The idea of the NRG is to obtain the quantum many-body states of
the system on all energy scales, or length scales, in a sequence of steps.
To do that, Wilson defined a basis of concentric wavefunctions in which the
Hamiltonian takes the form of a linear chain. A truncated chain with $n$
sites, described by an effective Hamiltonian $H_{n}$, gives the correct
physics on an energy scale $w_{n}$. The NRG transformation ${\bf T}$ relates
the Hamiltonians describing successive lower energy scales, $H_{n+1}={\bf T}%
[H_{n}]$, and leads to a systematic way of calculating the thermodynamic
properties at successive lower temperatures.

\begin{figure}[t]
\epsfxsize = 7.5cm
\begin{center}
\leavevmode
\epsffile{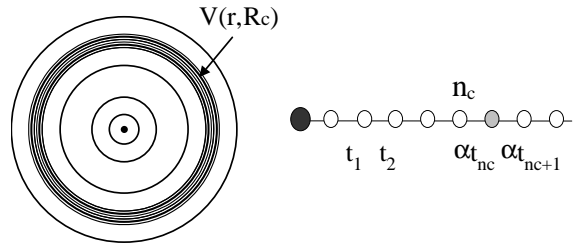}
\end{center}
\caption{Schematic representation of Wilson orbitals with the potential
barrier and the corresponding linear chain.}
\label{fig2}
\end{figure}

\begin{figure}[t]
\epsfxsize = 6.5cm
\begin{center}
\leavevmode
\epsffile{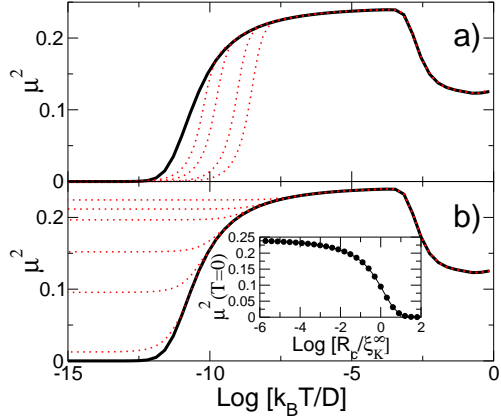}
\end{center}
\caption{(a) Impurity magnetic moment squared for a finite system with an
even number of particles as obtained with the NRG, the parameters are $D=1$, 
$U=0.02$, $\varepsilon _{d}=-0.01$, and $V=0.01$. The thick line corresponds
to the thermodynamic limit, thin lines, from right to left, to $%
n_{c}=46,\,50,\,56,$ and $60$. (b) Same as in a for (a) system with an odd
number of particles, the thin lines from top to bottom correspond to $%
n_{c}=45,\,51,\,55,\,61,\,65,$ and $71$. Inset: Zero temperature impurity
magnetic moment squared versus $\log (R_{c}/\xi _{K}^{\infty })$%
. }
\label{fig3}
\end{figure}

Going back to our problem and for the sake of simplicity, consider a
spherical metallic cluster of radius $R_{c}$ with the impurity at the
center. The cluster is embedded in a bulk material with which it is weakly
coupled through a large surface barrier. The Hamiltonian can be put in the
form $H=H_{AM}+V(r,R_{c})$ where the last term is a potential barrier placed
at a distance $R_{c}$ from the impurity. The Hamiltonian $H_{AM}$ is
rewritten in Wilson's basis as schematically shown in Fig.\ref{fig2}. The
potential barrier is described by including a higher diagonal energy to the $%
(n_{c}+1)$--shell centered around $R_{c}$. Alternatively the barrier can be
simulated with a smaller hopping matrix element in the region where $%
V(r,R_{c})$ is different from zero. We adopted this last description
reducing the hoppings by a factor $\alpha $. 
If the barrier is impenetrable ($\alpha =0$), the central cluster is
isolated. The process of renormalization ends up after $n_{c}$ steps and
with the obtained low energy states with a{\it \ definite number of particles%
} $N_{e}$ the partition function $Z$ is evaluated and the thermodynamic
properties of the system are obtained. This procedure shows that the high
temperature properties, evaluated with a number of shells $n<n_{c}$, the
impurity behaves as in a bulk material. Only when the temperature is lowered
and the discrete nature of the one-electron states becomes evident the
behavior of the system deviates from that of the macroscopic system. The
impurity susceptibility can be evaluated for even and odd number of
particles. As we discuss below, to describe even or odd number of particles
it is convenient to take even or odd number of shells $n_{c}$. The
susceptibility is calculated as:\cite{susc} 
\[
\chi ={\mu }_{i}^{2}(\sum_{\nu }\frac{P_{\nu }|\langle \nu |S_{iz}|\nu
\rangle |^{2}}{k_{B}T}+\sum_{\nu \neq \xi }\frac{2P_{\nu }|\langle \nu
|S_{iz}|\xi \rangle |^{2}}{E_{\xi }-E_{\nu }})
\]
where the summation is done over the low energy states $|\nu \rangle $ with
energies $E_{\nu }$ and $P_{\nu }=\exp (-E_{\nu }/k_{B}T)/Z$. The matrix
elements $\langle \nu |S_{iz}|\xi \rangle $ have to be evaluated in a
recursive way at each renormalization step. The susceptibility reflects the
thermodynamic properties of the system and we use it as an indicator of the
degree of screening of the impurity spin. In Fig.\ref{fig3} the impurity
susceptibility is shown for different system sizes. For comparison the same
quantity evaluated in the thermodynamic limit ($n_{c}\rightarrow \infty $)
is shown. As $n_{c}$ decreases the characteristic energy separation between
the one--electron states $\Delta $ increases. In the NRG this energy
separation at the Fermi level is $\Delta \simeq D\Lambda ^{-n_{c}/2}$ where $%
D$ is half the band width and $\Lambda \simeq 2$ is Wilson's discretization
parameter.\cite{Wilson} 
In Fig.\ref{fig3}(a) the effective impurity magnetic moment squared\cite
{Wilson} $\mu ^{2}=k_{B}T\chi $ for an even number of electrons is shown as
a function of the temperature. When the temperature is of the order of $%
\Delta $ the behavior deviates from that of the macroscopic sample and $\mu
^{2}$ decreases exponentially due to the gaps induced by confinement. For an
odd number of electrons the behavior is illustrated in Fig.\ref{fig3}(b). In
this case, as $k_{B}T$ approaches $\Delta $ the screening of the impurity
spin ends up and a finite magnetic moment survives down to zero temperature,
the ground state is a spin-1/2 doublet. In the inset the low temperature
magnetic moment versus $\log (R_{c}/\xi _{K}^{\infty })$ is shown with $\xi
_{K}^{\infty }$ being the Kondo screening length (see below), the behavior
shows the same tendency of that of Fig.\ref{fig1}(b) corresponding to exact
results of the impurity in a linear chain. Here the crossover between an
almost unscreened impurity spin to an almost completely screened spin occurs
for $\Delta \simeq $ $k_{B}T_{K}^{\infty }$.

The behavior of the system with a finite barrier is richer and more
interesting. Now the central cluster is in contact with a reservoir and the
number of electrons in the cluster is not a good quantum number. In the
numerical NRG approach the Fermi energy is set to zero. The one particle
spectrum in the absence of the impurity has a state at zero energy for even
iterations $n$ and two states at approximately $\pm D\Lambda ^{-n/2}$ for
odd $n$. Then by taking the barrier enclosing an even or odd number of
shells, $n_{c}$, we describe a system with the Fermi level at the energy of
a resonant state (at resonance)\ or between two of them (off resonance)
respectively. These two situations correspond approximately to an even and
odd  {\it average number} of electrons in the cluster respectively.

\begin{figure}[t]
\epsfxsize = 6.5cm
\begin{center}
\leavevmode
\epsffile{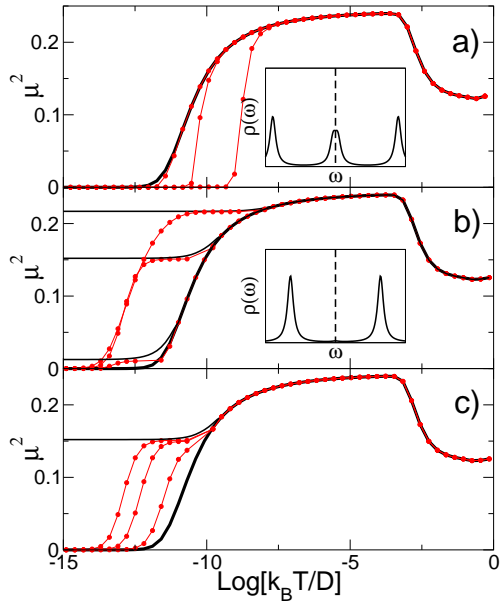}
\end{center}
\caption{ Impurity magnetic moment squared for a system with a finite
potential barrier and the Fermi energy at resonance (a) and off resonance
(b) and (c). Parameters are the same as in Fig. (3). In (a) and (b), $%
\alpha=0.05$ and different cluster sizes are shown ($n_{c}=50,\,62,\,72$ in
(a) and $n_{c}=49,\,61,\,71$ in (b)). In (c) the potential barrier is
changed for a fixed size: $n_{c}=61$, $\alpha =0,\,0.05,\,0.1,\,0.3,$ and $1$%
, (from left to right). In (b) and (c) the thin lines that extrapolate to a
finite value correspond to the isolated system. In a) and b) the insets show a
typical local density of states of the conduction electrons at the impurity site.}
\label{fig4}
\end{figure}

In Fig.\ref{fig4}(a) the impurity magnetic moment ${\mu }^{2}$ is shown for
the a system at resonance. Only when the temperature is lowered and reaches
the value $k_{B}T\approx \Delta $ the structure in the local density of
states becomes important. At low temperatures the universal Kondo behavior
is recovered with a Kondo temperature $T_{K}^{R}$ higher than $T_{K}^{\infty
}$, corresponding to a higher density of states. The crossover between the
high and low temperature regimes occurs in a small temperature interval
leading to a rapid decrease of the magnetic moment ${\mu }^{2}$.

For the system off resonance, as $k_{B}T$ approaches $\Delta $ the magnetic
moment saturates leading to a plateau in the curve ${\mu }^{2}$ versus $T$
as shown in Fig.\ref{fig4}(b). Only at lower temperatures the universal
Kondo behavior is recovered with a Kondo temperature $T_{K}^{OR}$ that is
lower than $T_{K}^{\infty }$. The plateau corresponds to a new regime of
partially screened magnetic moment. In this regime, as the temperature is
lowered, the thermodynamic properties do not change and are similar to those
of a finite system with an even number of particles. The central cluster has
a magnetic moment and it is the whole cluster that behaves as a Kondo
impurity or a ``Kondo grain'' when the temperature is lowered. For a given
potential barrier height the size of the grain can be changed: systems with
small $R_{c}$, a large $\Delta $, present a wide plateau with a
partially-screened magnetic moment ${\mu }_{ps}^{2}\sim 1/4$. As $R_{c}$
increases, both the width $\delta T$ of the plateau where the
partially-screened magnetic moment ${\mu }_{ps}$ is stable and the value of $%
{\mu }_{ps}^{2}$ decrease. For a given size, the $\delta T$ increases with
the barrier height as illustrated in Fig.\ref{fig4}(c).

In summary, for $k_{B}T_{K}^{\infty }\gg \Delta $ the fine structure (on the
scale of $T_{K}^{\infty }$) of the density of states does not change the
properties of the system and the universal Kondo behavior is recovered. For $%
k_{B}T_{K}^{\infty }\ll \Delta $ the local density of states, on the scale
of $T_{K}^{\infty }$, is a smooth function and the universal Kondo behavior
is obtained at low temperatures. In this case the system at resonance has a
Kondo temperature $T_{K}^{R}$ much larger than the corresponding to the
system out of resonance $T_{K}^{OR}$, however in both cases the high
temperature susceptibility is non universal. For $k_{B}T_{K}^{\infty }\sim
\Delta $ new confinement induced regimes are observed: for the system at
resonance, as the temperature is lowered, there is a rapid decrease in the
magnetic moment ${\mu }^{2}$ when $k_{B}T\sim \Delta $; for the system off
resonance there is a new partially-screened regime that leads to a plateau
in the temperature dependence of ${\mu }^{2}$. This plateau can be
interpreted as the free-spin regime of the whole Kondo grain that has
magnetic moment partially localized at the impurity. At very low
temperatures the electrons that are outside the cluster complete the
screening. The temperature at which this occurs depends on the coupling
between nanoscopic system and the bath. The condition for the existence of
these new regimes $k_{B}T_{K}^{\infty }\sim \Delta $ can be put in the form $%
\xi _{K}^{\infty }\sim R_{c}$ where $\xi _{K}^{\infty }=\hbar
v_{F}/k_{B}T_{K}^{\infty }$ is the Kondo screening length.

For the case of Co impurities in quantum corrals built on the Cu (111)
surface, taking $T_{K}^{\infty }=50K$ and using the parameters of the Cu
(111) surface band we estimate $\xi _{K}^{\infty }\simeq 400\AA $ which is
larger than the typical corral radius. Consequently if the Fermi level is
out of resonance we expect the Kondo screening to occur only at very low
temperatures. Larger corrals could be built or the effective barrier height
decreased by removing some atoms from the corral fence and in this way
control the effective Kondo temperature of the impurity.

Our results may have important consequences for the interpretation of
experiments with QD in Aharonov-Bohm (AB) rings.\cite{QD,Lee} In the AB
configuration, the local density of states inside the ring has the structure
of resonant states. A direct consequence of this structure are the
oscillations in the transmittance of a ring as a magnetic field threading
the circuit is varied. For rings built on GaAs as in some of the
experimental setups the energy separation between resonances $\Delta $ could
be larger than one Kelvin, in fact at 1K oscillations are still observed
indicating that at this temperature thermodynamic fluctuations do not wash
out the density of states structure. If the Kondo temperature of a quantum
dot inserted in one of the ring arms is of the order or smaller than $\Delta 
$, new effects are to be taken into account. In this case the quantum dot
behaves quite differently if the Fermi level lies at resonance or off
resonance, corresponding to a maximum or a minimum in the transmittance
respectively. The magnetic field can continuously shift the resonances and
if the system is in the temperature interval $T_{K}^{OR}$ 
\mbox{$<$}%
$T$ 
\mbox{$<$}%
$T_{K}^{R}$ , even at constant temperature the QD continuously evolves from
a low temperature regime ($T$ 
\mbox{$<$}%
$T_{K}^{R}$) to a high temperature regime ($T_{K}^{OR}$ 
\mbox{$<$}%
$T$ ). Moreover, the energy of the resonances depend not only on the
external field but also on the phase shift introduced by the QD that is
selfconsistently adjusted. On the light of our analysis, in this regime the
AB oscillations can not be interpreted directly in terms of a Kondo impurity
at a fixed temperature ratio $T/T_{K}$.

Acknowledgment: We thank I. Affleck and N. Andrei for stimulating
discussions. This work was partially supported by the CONICET and ANPCYT,
grants N. 02151 and 99 3-6343.


\begin{references}
\bibitem{susc}  The usual way of calculating the susceptibility in the NRG
is not appropriate for finite systems.
\end{references}
\end{document}